\documentstyle[12pt]{article}

\begin{document}
\begin{titlepage}

\centerline{\bf ON THE SELECTION OF TRIADS}
\centerline{\bf  IN THE TELEPARALLEL GEOMETRY}
\centerline{\bf AND BONDI'S RADIATING METRIC}
\bigskip
\centerline{\it K. H. Castello-Branco and J. W. Maluf$\,^{*}$}
\centerline{\it Instituto de F\'isica, Universidade de Bras\'ilia}

\centerline{\it C.P. 04385}
\centerline{\it 70.919-970  Bras\'ilia, DF}  
\centerline{\it Brazil}
\date{}
\begin{abstract}
A consistent Hamiltonian formulation of the teleparallel
equivalent of general relativity (TEGR) requires the theory
to be invariant under the global SO(3) symmetry group, which
acts on orthonormal triads in  three-dimensional spacelike
hypersurfaces. In the TEGR it is possible to make definite
statements about the energy of the gravitational field. In
this geometrical framework two sets of triads related by a
local SO(3) transformation yield different descriptions of
the gravitational energy. Here we consider the problem of
assigning  a unique set of triads to the metric tensor
restricted to the three-dimensional hypersurface. The
analysis is carried out in the context of Bondi's
radiating metric. A simple and original expression for
Bondi's news function is obtained, which allows us to
carry out numerical calculations and verify that 
a triad with a specific asymptotic behaviour yields the
minimum gravitational energy for a fixed space volume
containing the radiating source. This result supports the
conjecture that the requirement of a minimum gravitational
energy for a given space volume singles out uniquely the
correct set of orthonormal triads.

\end{abstract}
\thispagestyle{empty}
\vfill
\par
\noindent (*) e-mail: wadih@fis.unb.br
\end{titlepage}
\newpage

\noindent {\bf I. Introduction}\par
\bigskip
\noindent General covariance and the notion of absolute
parallelism of vector fields in space-time are the two major
geometrical concepts of teleparallel theories of gravity.
M\o ller\cite{Moller} was probably the first to make use of
the concept of absolute parallelism to put forward
an alternative geometrical framework for general relativity
that could satisfactorily address the problem of the
definition of the gravitational energy.

The idea of absolute parallelism can be  established by
considering a space-time vector field $V^\mu(x)$ and a
set of orthonormal tetrad fields $e^a\,_\mu(x)$. At the
space-time point $x^\lambda$ the tetrad components of the
vector field are given by $V^a(x)=e^a\,_\mu(x)V^\mu(x)$,
and at $x^\lambda +dx^\lambda$
by $V^a(x+dx)=e^a\,_\mu(x+dx)V^\mu(x+dx)=V^a(x)+DV^a(x)$,
where $DV^a(x)=e^a\,_\mu(\nabla_\lambda V^\mu)dx^\lambda$.
The covariant derivative $\nabla$ is constructed out of the
connection

$$\Gamma^\lambda_{\mu\nu}=
e^{a\lambda}\partial_\mu e_{a\nu}\;.\eqno(1)$$

\noindent The vector field $V^\mu(x)$ is said to be autoparallel
if its tetrad components at distant points coincide. Thus
$V^\mu(x)$ is autoparallel if $\nabla_\lambda V^\mu$ vanishes.
Therefore connection (1) defines a {\it condition} for absolute
parallelism, or teleparallelism, in space-time. Such connection
only makes sense if the tetrad field transforms
under the {\it global} SO(3,1) group.

A gravity theory based on (1) obviously depart from the
Riemannian geometry because the curvature tensor constructed
out of it vanishes identically. Inspite of this fact, there
does exist a theory based on (1) that describes the dynamics
of the gravitational field in agreement with Einstein's
general relativity. M\o ller called such theory the ``tetrad
theory of gravity", but for a long time it has been known as
the teleparallel equivalent of general relativity
(TEGR)\cite{Hehl1,Hehl2}. The Lagrangian density for
this alternative description of general relativity
is constructed by means of a quadratic combination of the
torsion tensor $T^a\,_{\mu \nu}=\partial_\mu e^a\,_\nu-
\partial_\nu e^a\,_\mu$, which is related to the antisymmetric
part of connection (1).

It is possible to establish a theory for the gravitational
field directly from (1). However, in order to make contact with
more recent analysis,  it is instructive to approach
the TEGR firstly considering it with a local SO(3,1) symmetry
in the Lagrangian context. Eventually we will return to the
geometrical framework determined by (1).

Although not extensively investigated in the
literature, the TEGR in Lagrangian form has been
considered as a viable formulation of the gravitational
dynamics\cite{Shirafuji,Baekler,Nester,Mielke,Pereira} 
inspite of troubles that may spoil the initial value
problem\cite{Kop,Muller}. Such problems take place if the
Lagrangian density of the TEGR is formulated with a {\it local}
SO(3,1) symmetry, in which case the torsion tensor is defined by
$T^a\,_{\mu\nu}=\partial_\mu e^a\,_\nu-\partial_\nu e^a\,_\mu+
\omega_\mu\,^a\,_b\, e^b\,_\nu-\omega_\nu\,^a\,_b\, e^b\,_\mu$.
The spin connection $\omega_{\mu ab}$ is totally independent
of $e^a\,_\mu$ and satisfies the condition of vanishing curvature:
$R^a\,_{b\mu\nu}(\omega)\equiv 0$.

In order to solve problems with respect to the initial value
problem, the Hamiltonian formulation of the TEGR with the local
SO(3,1) symmetry was considered\cite{Maluf1}. By working out
the constraint algebra  one ultimately concludes that in order
to formulate consistently the theory in terms of {\it first
class constraints} it is mandatory to break the local SO(3)
symmetry of the action in Hamiltonian form.
Therefore a consistent Hamiltonian formulation
of the theory displays invariance under the global SO(3) group
that acts on triads restricted to 
three-dimensional spacelike hypersurfaces.

The major feature of the Hamiltonian formulation of the TEGR
is that the integral form of the Hamiltonian constraint
equation can be written as an energy equation of the type
$C=H-E=0$\cite{Maluf2,GRG1,GRG2}. The
reason for this follows from the fact that the Hamiltonian
constraint contains a scalar density in the form of a total
divergence, whose integral over the whole three-dimensional
hypersurface yields the ADM energy\cite{ADM}. As we will
argue, the integral of this scalar density over finite
volumes of the three-dimensional hypersurface yields a
natural definition for the gravitational field energy. 
Therefore in the TEGR one can make definite statements about
the localizability of the gravitational energy, inspite of
claims according to which the latter is not localizable.
In fact the very concept of a black hole lends support to
the idea that the gravitational energy is localizable, since
there is no process by means of which the gravitational mass
inside a black hole can be made to vanish.  A remarkable
application of the energy expression of the TEGR has been
made in the evaluation of the irreducible mass of a rotating
black hole\cite{Maluf3}.

The TEGR exhibits two specific properties: the
emergence of a possible definition for the gravitational field
energy and the global SO(3) symmetry of the theory. We believe
that these two features are intimately related. Since the
symmetry of the theory is global, triads related by a local
SO(3) transformation are inequivalent and 
a priori we have no means to select the one
that actually describes the spacelike hypersurface. We
conjecture that the requirement of a minimum gravitational
energy for a given space volume is one condition 
that singles out uniquely the correct set of triads. In this
paper we investigate this conjecture in the realistic context
of Bondi's radiating metric\cite{Bondi}. This conjecture was
already put forward in a previous investigation of Bondi's
energy in the framework of the TEGR\cite{Maluf4}. An
additional, essential requirement for a consistent
expression of the gravitational energy is the boundary
conditions on the triads. It was noted\cite{Maluf2} that the
ADM energy is obtained from the energy expression
of the TEGR if the asymptotic behaviour of the triads at
spacelike infinity is given by

$$e_{(i)j}\approx \eta_{ij} +
{1\over 2}h_{ij}({1\over r})\;,\eqno(2)$$

\noindent irrespective of any symmetry of the tensor $h_{ij}$.

A second possible, independent condition for assigning a set
of triads to a given three-dimensional metric tensor amounts
to requiring a symmetric tensor $h_{ij}=h_{ji}$ in the
asymptotic expansion of $e_{(i)j}$. We will prove that
this symmetry condition {\it uniquely} associates $e_{(i)j}$ 
to the metric tensor of the spacelike hypersurface of
Bondi's metric. The unique character of such triads 
strongly supports this second conjecture.

The two conjectures above are not mutually excluding. In
this paper we argue that the set
of triads that yield the minimum gravitational energy
within a space volume containing the radiating source is
the one whose asymptotic behaviour is determined by the
symmetry condition $h_{ij}=h_{ji}$. We will show that this
fact is indeed verified by analyzing several
configurations for the triads. Unfortunately we have
not found it possible to prove on general grounds that the
``symmetrized" triad yields the minimum energy.

In order to obtain numerical values for the gravitational
energy for a finite three-dimensional volume of Bondi's
space-time we need an explicit expression of the news function.
However the existing expressions in the literature are not
suitable for our purposes. In particular, the expression of
the news function given by Hobill\cite{Hobill} (to be
presented ahead) is very intricate to the  extent of not
allowing a computer evaluation of numerical values of the
gravitational energy. Therefore we have obtained an original
and simpler expression for the news function that: (i)
satisfies all necessary regularity conditions related to the
axial symmetry of the system, (ii) makes the initial and final
states (space-times) described by Bondi's metric to be
nonradiative and (iii) yields an expression for the energy
density that can be numerically integrated.

In Section II we describe the TEGR in Lagrangian and 
Hamiltonian formulations, show the emergence of the
definition of the gravitional energy and further discuss
the troubles that arise in the initial value problem of the
theory if it is constructed with a local SO(3,1) symmetry.
In section III we present Bondi's radiating metric and three
expressions for triads restricted to the three-dimensional
spacelike hypersurface. In this section we also prove that the
symmetry condition $h_{ij}=h_{ji}$ uniquely associates a set
of triads (whose asymptotic behaviour is given by (2)) with the
metric tensor for the spacelike section of asymptotically flat
space-times. The news function and the
related mass aspect that will be needed for the calculations of
the gravitational energy are obtained in section IV.
In section V we carry out several calculations that lead to
the main conclusion regarding the selection of triads.\par
\bigskip

\noindent Notation: spacetime indices $\mu, \nu, ...$ and local Lorentz 
indices $a, b, ...$ run from 0 to 3. In the 3+1 decomposition latin 
indices from the middle of the alphabet indicate space indices according 
to $\mu=0,i,\;\;a=(0),(i)$. The tetrad field $e^a\,_\mu$ and
the spin connection $\omega_{\mu ab}$ yield the usual definitions
of the torsion and curvature tensors:  $R^a\,_{b \mu \nu}=
\partial_\mu \omega_\nu\,^a\,_b +
\omega_\mu\,^a\,_c\,\omega_\nu\,^c\,_b\,-\,...$,
$T^a\,_{\mu \nu}=\partial_\mu e^a\,_\nu+
\omega_\mu\,^a\,_b\,e^b\,_\nu\,-\,...$. The flat space-time metric 
is fixed by $\eta_{(0)(0)}=-1$.\par
\bigskip
\bigskip
\bigskip
\noindent {\bf II. The Lagrangian and Hamiltonian formulations
of the TEGR}\par
\bigskip

The Lagrangian density of the TEGR in empty space-time, displaying
a local SO(3,1) symmetry, is given by

$$L(e,\omega,\lambda)\;=\;-ke({1\over 4}T^{abc}T_{abc}\,+\,
{1\over 2}T^{abc}T_{bac}\,-\,T^aT_a)\;+\;
e\lambda^{ab\mu\nu}R_{ab\mu\nu}(\omega)\;.\eqno(3)$$

\noindent where $k={1\over {16\pi G}}$, $G$ is the gravitational 
constant; $e\,=\,det(e^a\,_\mu)$, $\lambda^{ab\mu\nu}$ are 
Lagrange multipliers and $T_a$ is the trace of the torsion tensor
defined by $T_a=T^b\,_{ba}$.   The tetrad field $e_{a\mu}$ and the
spin connection $\omega_{\mu ab}$ are completely independent field
variables.  The latter is enforced to satisfy the condition of
zero curvature. Therefore this Lagrangian formulation is in no way
similar to the usual Palatini formulation, in which the spin
connection is related to the tetrad field via field equations.
Later on we will introduce the  tensor $\Sigma_{abc}$ defined by

$${1\over 4}T^{abc}T_{abc} + {1\over 2}T^{abc}T_{bac}-T^aT_a\;\equiv\;
T^{abc}\Sigma_{abc}\;.$$

The equivalence of the TEGR with Einstein's general relativity is         
based on the identity

$$eR(e,\omega)\;=\;eR(e)\,+\,
e({1\over 4}T^{abc}T_{abc}\,+\,{1\over 2}T^{abc}T_{bac}\,-\,T^aT_a)\,-\,
2\partial_\mu(eT^{\mu})\;,$$

\noindent which is obtained by just substituting the arbitrary
spin connection $\omega_{\mu ab}\,=\,^o\omega_{\mu ab}(e)\,+\,
K_{\mu ab}$ in the scalar curvature tensor $R(e,\omega)$ in the
left hand side; $^o\omega_{\mu ab}(e)$ is the Levi-Civita 
connection and $K_{\mu ab}\,=\,
{1\over 2}e_a\,^\lambda e_b\,^\nu(T_{\lambda \mu \nu}+
T_{\nu \lambda \mu}-T_{\mu \nu \lambda})$ is the contorsion tensor.
The vanishing of $R^a\,_{b\mu\nu}(\omega)$, which is one of the
field equations derived from (3), implies the equivalence of 
the scalar curvature $R(e)$, constructed out of $e^a\,_\mu$ only, 
and the quadratic combination of the torsion tensor. It also
ensures that the field equation arising from the variation of
$L$ with respect to $e^a\,_\mu$ is strictly equivalent to
Einstein's equations in tetrad form. Let 
${{\delta L}\over {\delta {e^{a\mu}}}}=0$ denote the field equations
satisfied by $e^{a\mu}$. It can be shown by explicit calculations
that

$${{\delta L}\over {\delta {e^{a\mu}}}}\;=\;
{1\over 2}e \lbrace R_{a\mu}(e)\,
-\,{1\over 2}e_{a\mu}R(e)\rbrace\;.\eqno(4)$$

\noindent We refer the reader to
Ref. \cite{Maluf1} for additional details.

Throughout this section we will be interested in asymptoticaly 
flat space-times. The Hamiltonian formulation of the TEGR can be 
successfully implemented if we fix the gauge $\omega_{0ab}=0$ from 
the outset, since in this case the constraints (to be 
shown below) constitute a {\it first class} set\cite{Maluf1}.
The condition $\omega_{0ab}=0$ is achieved by breaking the local
Lorentz symmetry of (3). We still make use of the residual time
independent gauge symmetry to fix the usual time gauge condition
$e_{(k)}\,^0\,=\,e_{(0)i}\,=\,0$. Because of $\omega_{0ab}=0$,
$H$ does not depend on $P^{kab}$, the momentum canonically 
conjugated to $\omega_{kab}$. Therefore arbitrary variations of
$L=p\dot q -H$ with respect to $P^{kab}$ yields 
$\dot \omega_{kab}=0$. Thus in view of $\omega_{0ab}=0$, 
$\omega_{kab}$ drops out from our considerations.

As a consequence of the above gauge fixing the canonical action 
integral obtained from (3) becomes\cite{Maluf1}

$$A_{TL}\;=\;\int d^4x\lbrace \Pi^{(j)k}\dot e_{(j)k}\,-\,H\rbrace\;,
\eqno(5)$$

$$H\;=\;NC\,+\,N^iC_i\,+\,\Sigma_{mn}\Pi^{mn}\,+\,
{1\over {8\pi G}}\partial_k (NeT^k)\,+\,
\partial_k (\Pi^{jk}N_j)\;.\eqno(6)$$

\noindent $N$ and $N^i$ are the lapse and shift functions, and 
$\Sigma_{mn}=-\Sigma_{nm}$ are Lagrange multipliers. The
constraints are defined by 

$$ C\;=\;\partial_j(2keT^j)\,-\,ke\Sigma^{kij}T_{kij}\,-\,
{1\over {4ke}}(\Pi^{ij}\Pi_{ji}-{1\over 2}\Pi^2)\;,\eqno(7a)$$

$$C_k\;=\;-e_{(j)k}\partial_i\Pi^{(j)i}\,-\,
\Pi^{(j)i}T_{(j)ik}\;,\eqno(7b)$$

\noindent where $e=det(e_{(j)k})$, $T^i\,=\,g^{ik}e^{(j)l}T_{(j)lk}$,
$T_{(i)jk}=\partial_j e_{(i)k}-\partial_k e_{(i)j}$, and

$$\Sigma^{ijk}={1\over 4}(T^{ijk}+T^{jik}-T^{kij})+
{1\over 2}(\eta^{ik}T^j-\eta^{ij}T^k)\;.$$

We remark that (5) and (6) are now invariant under
global SO(3) and general coordinate transformations. Therefore
the torsion tensor restricted to the three-dimensional spacelike
hypersurface is ultimately related to the antisymmetric part
of the spatial components of (1). Had we dispensed with the
connection $\omega_{\mu ab}$ from the outset we would have
arrived at precisely the canonical formulation determined by
(5), (6) and (7).
Such connection has been considered in previous
investigations of teleparallel theories, but in fact it is
eventually unnecessary for the establishment of the theory.

If we assume the asymptotic behaviour 

$$e_{(j)k}\approx \eta_{jk}+
{1\over 2}h_{jk}({1\over r})\eqno(2)$$ 

\noindent for $r \rightarrow \infty $, then in view of 
the relation

$${1\over {8\pi G}}\int d^3x\partial_j(eT^j)\;=\;
{1\over {16\pi G}}\int_S dS_k(\partial_ih_{ik}-\partial_kh_{ii})
\; \equiv \; E_{ADM}\;\eqno(8a)$$

\noindent where the surface integral is evaluated for 
$r \rightarrow \infty$, the integral form of 
the Hamiltonian constraint $C=0$ may be rewritten as

$$\int d^3x\biggl\{ ke\Sigma^{kij}T_{kij}+
{1\over {4ke}}(\Pi^{ij}\Pi_{ji}-{1\over 2}\Pi^2)\biggr\}
\;=\;E_{ADM}\;.\eqno(8b)$$

\noindent The integration is over the whole three-dimensional
space. Given that $\partial_j(eT^j)$ is a scalar  density,
from (8a,b) we define the gravitational
energy density enclosed by a volume V of the space as

$$E\;=\;{1\over {8\pi G}}\int_V d^3x\partial_j(eT^j)\;.\eqno(9)$$  

\noindent It must be noted that $E$ depends only on the triads
$e_{(k)i}$ restricted to a three-dimensional spacelike
hypersurface; the inverse quantities $e^{(k)i}$ can be written
in terms of $e_{(k)i}$. From the right hand side of 
equation (4) we observe that the dynamics of
the triads does not depend on $\omega_{\mu ab}$. Therefore $E$
given above does not depend on the fixation of any gauge for
$\omega_{\mu ab}$. The reference space which defines the zero
of gravitational energy has been defined in ref.\cite{Maluf3}.
We briefly remark that the differences between (9) and
M\o ller's expression for the gravitational energy have been
thoroughly discussed in ref.\cite{Maluf4}.

We make now the important assumption that the general form of
the canonical structure of the TEGR is the same for any class
of space-times, irrespective of the peculiarities of the latter
(for the de Sitter space\cite{Maluf5}, for example, there is an
{\it additional} term in the Hamiltonian constraint $C$).
Therefore we assert that the integral form of the Hamiltonian 
constraint equation can be written as $C=H-E=0$ for any 
space-time, and that (9) represents the gravitational 
energy for arbitrary space-times with any topology.

We recall finally that
M\"uller-Hoissen and Nitsch\cite{Muller} and 
Kopczy\'nski\cite{Kop} have shown that in general
the theory defined by (3) faces difficulties with respect to the 
Cauchy problem. They have shown that in general six components of
the torsion tensor are not determined from the evolution of the 
initial data. On the other hand, the constraints of the  theory
constitute a first class set provided we fix the six quantities
$\omega_{0ab}=0$ {\it before varying the action}\cite{Maluf1}. 
This condition 
is mandatory and does not merely represent one particular gauge
fixing of the theory. Since the fixing of $\omega_{0ab}$ yields a
well defined theory with first class constraints, we cannot 
assert that the field configurations of the
latter are gauge equivalent to configurations
whose time evolution is not precisely determined. The
requirement of local SO(3,1) symmetry plus the addition of 
$\lambda^{ab\mu \nu}R_{ab\mu\nu}(\omega)$ in (3) has the
ultimate effect of discarding the connection.

Constant rotations constitute a  basic feature of the teleparallel   
geometry. According to M\o ller\cite{Moller}, in the framework
of the abolute parallelism tetrad fields, together with the
boundary conditions, uniquely determine a {\it tetrad lattice}, 
apart from an arbitrary {\it constant rotation of the tetrads in the 
lattice}.  \par

\bigskip
\bigskip
\bigskip
\noindent {\bf III. Bondi's radiating metric and the
associated triads}.\par

\bigskip
Bondi's metric describes the asymptotic form of a radiating
solution of Einstein's equations. It is not an exact solution;
it holds only in the asymptotic region. In terms of
radiation coordinates $(u,r,\theta,\phi)$, where $u$ is the
retarded time and $r$ is the luminosity distance, Bondi's
metric is written as\cite{Bondi}

$$ds^2\;=\;-\biggl( {V\over r} e^{2\beta}-
U^2\,r^2 e^{2\gamma}\biggr)du^2
-2e^{2\beta}du\,dr - 2U\,r^2\,e^{2\gamma}du\,d\theta$$

$$+r^2 \biggl( e^{2\gamma}\,d\theta^2 +
e^{-2\gamma}\,sin^2\theta\,d\phi^2\biggr)\;.\eqno(10)$$

\noindent This metric tensor displays axial symmetry
and reflection invariance.
By requiring $u=constant$, (10) describes null hypersurfaces.
Each null radial (light) ray is labelled by particular
values of $u, \theta$ and $\phi$. At spacelike infinity $u$
takes the standard form $u=t-r$. The four quantities appearing
in (10), $V, U, \beta$ and $\gamma$ are functions of
$u, r$ and $\theta$. A more general form of this metric has
been given by Sachs\cite{Sachs}, who  showed
that the most general metric tensor describing asymptotically
flat gravitational waves depends on six functions of the
coordinates.

The functions in (10) have the following asymptotic
behaviour:

$$\beta\;=\;-{c^2\over {4r^2}}+...$$

$$\gamma\;=\;{ c \over r}+O({1\over {r^3}})+...$$

$${V\over r}\;=\;1\,-\,{{2M}\over r}\,
-\,{1\over r^2}\biggl[
{{\partial d}\over {\partial \theta}} +
d\,cot\theta-
\biggl({{\partial c}\over{\partial \theta}}\biggr)^2
-4c\biggl({{\partial c}\over
{\partial \theta}}\biggr)cot\theta-
{1\over 2}\,c^2\biggl(1+8cot^2\theta\biggr)\biggr]+...$$

$$U\;=-\;{1\over r^2}\biggl(
{{\partial c}\over {\partial \theta}}
+2c\,cot\theta\biggr)+ 
{1\over{r^3}}\biggl(2d+3c\,
{{\partial c}\over{\partial \theta}}
+4c^2\,cot\theta\biggr)+...$$

\noindent where $M=M(u,\theta)$ and $d=d(u,\theta)$ are the
mass aspect and the dipole aspect, respectively. From the
function $c(u,\theta)$ we define the {\it news function}
${{\partial c(u,\theta)} \over {\partial u}}$.

The functions $U, V, \beta$ and $\gamma$ must satisfy
{\it regularity conditions} along the $z$ axis
($\theta=0,\pi$). We must require

$$V,\;\;\;\beta,\;\;\;{U\over {sin\theta}},\;\;\;
{\gamma\over{sin^2\theta}}$$

\noindent to be regular functions of $cos\theta$ for
$\theta=0,\pi$. The regularity conditions will be necessary
for the construction of the news function, in section IV.

The application of (9) to Bondi's metric requires transforming
it to spherical coordinates ($t, r, \theta, \phi$) for which
$t=constant$ defines a space-like hypersurface.  Therefore
we carry out a coordinate transformation such that the new
timelike coordinate is given by $t=u+r$. We arrive at

$$ds^2\;=\; -\biggl({V\over r}e^{2\beta}-
U^2\,r^2\,e^{2\gamma}\biggr)dt^2
-2U\,r^2\,e^{2\gamma}dt\,d\theta$$

$$+2\biggl[e^{2\beta}\biggl({V\over r}-1\biggr)-
U^2\,r^2\,e^{2\gamma}\biggr]dr\,dt$$

$$+\biggl[e^{2\beta}\biggl(2-{V\over r}\biggr)+
U^2\,r^2\,e^{2\gamma}\biggr]dr^2
+2U\,r^2\,e^{2\gamma}dr\,d\theta + 
r^2\biggl( e^{2\gamma}d\theta^2+
e^{-2\gamma}\,sin^2\theta\,d\phi^2\biggr)\;.\eqno(11)$$

\noindent Therefore the metric restricted to a
three-dimensional spacelike hypersurface is given by

$$ds^2\;=\;\biggl[ e^{2\beta}\biggl(2-{V\over r}\biggr)+
U^2\,r^2\,e^{2\gamma}\biggr] dr^2+
2U\,r^2\,e^{2\gamma}\,dr\,d\theta$$

$$+r^2\biggl(e^{2\gamma}\,d\theta^2+
e^{-2\gamma}\,sin^2\theta\,d\phi^2\biggr)\;.\eqno(12)$$

\noindent We recall that Goldberg\cite{Goldberg} and
Papapetrou\cite{Papapetrou1} have already considered Bondi's
metric in cartesian coordinates.

The crucial point of the present investigation is the
determination, in the framework of the TEGR, of
the correct set of
triads that lead to (12). If the metric tensor has only
diagonal components,
such as the metric tensor restricted to the three-dimensional
spacelike section of Kerr's space-time, then the {\it simplest}
construction has proven to be the correct one\cite{Maluf3}.
However, for metric tensors that contain off-diagonal
terms, the determination of the unique triad is by no means
a trivial procedure. In fact there is an infinity of triads
that satisfy the boundary conditions determined by (2) and
lead to (12). In Ref. \cite{Maluf4} two sets of triads that
comply with (2) are presented. They are given by

$$e_{(k)i}\;=\;\pmatrix{ A\,sin\theta\,cos\phi
+B\,cos\theta\,cos\phi&
rC\,cos\theta\,cos\phi & -rD\,sin\theta\,sin\phi\cr
A\,sin\theta\,sin\phi+B\,cos\theta\,sin\phi &
rC\,cos\theta\,sin\phi & rD\,sin\theta\,cos\phi\cr
A\,cos\theta - B\,sin\theta &
-rC\,sin\theta & 0\cr}\;,\eqno(13)$$

\noindent where

$$A\;=\;e^\beta \sqrt{2-{V\over r}}\;,\eqno(14a)$$

$$B\;=\;r\,U\,e^\gamma\;,\eqno(14b)$$

$$C\;=\;e^\gamma\;,\eqno(14c)$$

$$D\;=\;e^{-\gamma}\;,\eqno(14d)$$

\noindent and

$$e_{(k)i}\;=\;\pmatrix{ A'\,sin\theta\,cos\phi&
rB'\,cos\theta\,cos\phi+rC'\,sin\theta\,cos\phi&
-rD'\,sin\theta\,sin\phi\cr
A'\,sin\theta\,sin\phi&rB'\,cos\theta\,sin\phi
+rC'\,sin\theta\,sin\phi&
rD'\,sin\theta\,cos\phi\cr
A'\,cos\theta&-rB'\,sin\theta +
rC'\,cos\theta&0\cr}\;,\eqno(15)$$

\noindent where

$$A'\;=\;\bigg[ e^{2\beta}\biggl( 2-{V\over r}\biggr)+
U^2\,r^2\,e^{2\gamma}\biggr]^{1\over 2}\;,\eqno(16a)$$

$$B'\;=\;{1\over A'}\,e^{\beta +
\gamma}\sqrt{2-{V\over r}}\;,\eqno(16b)$$

$$C'\;=\;{1\over A'}\, U\,r\,e^{2\gamma}\;,\eqno(16c)$$

$$D'\;=\;e^{-\gamma}\;.\eqno(16d)$$

\noindent It is easy to see that both (13) and (15)
yield the metric tensor (12) through the relation
$e_{(i)j}e_{(i)k}=g_{jk}$. They are related by a
{\it local} SO(3) transformation.

We have presented (13) and (15) because they
are the {\it simplest} constructions that satisfy
two basic requirements: ({\bf i}) the triads must have
the asymptotic behaviour given by (2); ({\bf ii}) by making
the physical parameters of the metric vanish we must have
$T_{(k)ij}=0$ everywhere. In the present case if we make
$M=d=c=0$ both (13) and (15) acquire the form

$$e_{(k)i}\;=\;\pmatrix{sin\theta\,cos\phi &
r\,cos\theta\,cos\phi &
-r\,sin\theta\,sin\phi\cr
sin\theta\,sin\phi &
r\,cos\theta\,sin\phi &
r\,sin\theta\,cos\phi\cr
cos\theta &
-r\,sin\theta & 0\cr}\;.\eqno(17)$$

\noindent In cartesian coordinates the expression above can
be reduced to the diagonal form $\;\;$
$e_{(k)i}(x,y,z)=\delta_{ik}$. The requirement (ii) above
is essentialy equivalent to the establishment of 
reference space triads, as discussed in \cite{Maluf3}.
A proper definition of gravitational energy
requires the notion of  reference space triads, which in
the present case are given by (17). Note
that by a suitable choice of a local SO(3) rotation we
can make the flat space triads (17) satisfy the
requirement (i), but not (ii).

We proceed now to obtain the triads whose aymptotic expansion
is given by (2) with the symmetry condition $h_{ij}=h_{ji}$.
The procedure is the following. We consider, for instance,
triads (13) and transform it to cartesian coordiates. Then we
perform a local, asymptotic transformation

$$\tilde e_{(k)i}(t,x,y,z)=
\Lambda^{(j)}\,_{(k)}\,e_{(j)i}(t,x,y,z)\;,\eqno(18)$$

\noindent where $\Lambda^{(j)}\,_{(k)}$ satisfies

$$\Lambda^{(j)}\,_{(k)}\approx \delta^{(j)}\,_{(k)}+
\omega^{(j)}\,_{(k)}\;,\eqno(19a) $$

$$\omega_{(j)(k)}=-\omega_{(k)(j)}\;,\eqno(19b)$$

\noindent and $\omega_{(j)(k)} \sim O({1\over r})$ for
$r\rightarrow \infty$ ($r=\sqrt{x^2+y^2+z^2}$).
Transformation (19) preserves the asymptotic behaviour of
the triads. By substituting (19) in (18) we find

$$\tilde e_{(k)i}=e_{(k)i}+\omega_{(k)i}\;,$$

\noindent from what follows

$$\tilde h_{ki}= h_{ki}+2\omega_{ki}\;,\eqno(20)$$

\noindent where $h_{ki}$ is given by the asymptotic expansion
of (13) in cartesian coordinates. By requiring

$$\tilde h_{ki}=\tilde h_{ik}\;,\eqno(21)$$

\noindent and making use of (19b) we find that

$$\omega_{ki}={1\over 4}(h_{ki}-h_{ik})\;.\eqno(22)$$

\noindent Substituting now (22) in (20) we arrive at

$$\tilde h_{ki}={1\over 2}(h_{ki}+h_{ik})
\equiv h_{(ki)}\;.\eqno(23)$$

\noindent Thus we ultimately obtain the symmetrized triads
in cartesian coordinates:

$$\tilde e_{(k)i} \approx
\eta_{ki}+{1\over 2}\tilde h_{ki}\;.\eqno(24)$$

It must be noted that we arrive at a symmetrized triad
irrespective of the triads we consider initially. The only
requirement is that the unrotated triad must satisfy
the asymptotic behaviour (2). Had we considered (15) we
would arrive at the same result. In particular, from
(22) we observe that if the initial triad is already
symmetrized, then no local, asymptotic rotation is
necessary.

By transforming (24) into spherical coordinates
$t,r,\theta,\phi$ (in which
case the triads are no longer symmetrized), we finally
arrive at

$$\tilde e_{(1)1}\approx (1+{M\over r})sin\theta\,cos\phi-
{f\over {2r}}cos\theta\,cos\phi\;,$$

$$\tilde e_{(2)1}\approx (1+{M\over r})sin\theta\,sin\phi-
{f\over {2r}}cos\theta\,sin\phi\;,$$

$$\tilde e_{(3)1}\approx (1+{M\over r})cos\theta+
{f\over {2r}}sin\theta\;,$$

$$\tilde e_{(1)2}\approx r(1+{c\over r})cos\theta\,cos\phi-
{f\over 2}sin\theta\,cos\phi\;,$$

$$\tilde e_{(2)2}\approx r(1+{c\over r})cos\theta\,sin\phi-
{f\over 2}sin\theta\,sin\phi\;,$$

$$\tilde e_{(3)2}\approx -r(1+{c\over r})sin\theta-
{f\over 2}cos\theta\;,$$

$$\tilde e_{(1)3}\approx
-r(1-{c\over r})sin\theta\,sin\phi\;,$$

$$\tilde e_{(2)3}\approx
r(1-{c\over r})sin\theta\,cos\phi\;,$$

$$\tilde e_{(3)3}\approx 0\;,\eqno(25)$$

\noindent where $f$ is given by

$$f={{\partial c}\over{\partial \theta}}+ 2c\,cotg\theta\;.$$

\noindent We observe that by making $M=c=0$ triads (25) reduce
to (17). We also note that (25), as well as (13) and (15), only
make sense in the asymptotic region where Bondi's metric is
valid.

In view of (23) and (24) it is now easy to prove 
the uniqueness of the symmetrized triad. Suppose that the
asymptotic behaviour of the metric tensor is given by

$$g_{ij}\approx \eta_{ij}+
h_{ij}^\star({1\over r})\;.\eqno(26)$$

\noindent  On the other hand by making use of (2) it follows
from the relation $g_{ij}=e^{(k)}\,_i e_{(k)j}$ and from
(23) that

$$ g_{ij}\approx \eta_{ij}+
{1\over 2}(h_{ij}+h_{ji})=\eta_{ij}+h_{(ij)}\;.\eqno(27)$$

\noindent Since $h_{ij}^\star$ is unique, by comparing (26)
and (27) we are led to conclude that there exists a unique
symmetrized triad associated to the spatial section of an
asymptotically flat metric tensor.\par

\bigskip
\bigskip
\noindent {\bf IV. Construction of the news function and of the
mass aspect}\par
\bigskip

In order to establish explicit expressions for the functions
$c(u,\theta)$ and $M(u,\theta)$, the suplementary field
equations $R_{00}=R_{02}=0$ for Bondi's metric 
are considered. In simplified form they are
given by\cite{Bondi}

$${{\partial M}\over{\partial u}}=-\biggl(
{{\partial c}\over {\partial u}}\biggr)^2+
{1\over 2}{\partial \over {\partial u}}\biggl[
{{\partial ^2 c}\over {\partial \theta^2}}+
3{{\partial c}\over {\partial\theta}}\,cotg \theta-
2c\biggr]\;,\eqno(28)$$

$$-3{{\partial d}\over {\partial u}}=
{{\partial M}\over{\partial \theta}}+
3c\,{{\partial^2 c}\over{\partial \theta\,\partial u}}
+4c\,{{\partial c}\over{\partial u}}\,cotg \theta
+{{\partial c}\over{\partial \theta}}
{{\partial c}\over{\partial u}}\;.\eqno(29)$$

\noindent From (28) we observe that if
${{\partial c}\over{\partial u}}=0$, the mass aspect $M$
does not depend on $u$.

For a family of null hypersurfaces Bondi's mass is defined by

$$m(u)={1\over 2}
\int_0^\pi M(u,\theta)\,sin\theta\,d\theta\;.\eqno(30)$$

\noindent It represents the mass of the system at the
retarded time $u$. Multiplying both sides of (28) by
$sin\theta$, integrating in $\theta$ and making use of
the regularity conditions on the $z$ axis stated in
section III, we arrive at

$${{d m}\over{d u}}=-{1\over 2}\int_0^\pi
\biggl({{\partial c}\over {\partial u}}\biggr)^2\,
sin\theta\,d\theta\;,\eqno(31)$$

\noindent which expresses the loss of mass. This equation
was first obtained
by Bondi et. al.\cite{Bondi}. Therefore if the news function
is nonvanishing, the mass of the system decreases in time.
We remark that not only $c(u,\theta)$, but also the news
function must be a regular function on the $z$ axis.

The function $c(u,\theta)$ determines not only the loss of
mass, but in fact it determines the whole structure of
Bondi's metric, since via (28) and (29) it also determines
the functions $M$, $d$, and all other functions that arise
in the process of integration of the field equations.
Therefore its construction deserves special attention.

The news function proposed by Bondi et. al.\cite{Bondi} is
given by

$${{\partial c}\over{\partial u}}=\sum_{n=0}^\infty
f_n(u)P_n(\mu)\;,$$

\noindent where $\mu=cos\theta\;$. $P_n(\mu)$ are the
Legendre polynomials and $f_n(u)$ are functions to be
determined. These functions become more and more intricate
for increasing $n$, and eventually the above expression
cannot be manipulated analytically.

Bonnor and Rotenberg\cite{Bonnor}, Papapetrou\cite{Papapetrou2}
and Hallidy and Janis\cite{Hallidy} have attempted at
establishing an expression for $c(u,\theta)$, but none of
these proposals worked out satisfactorily, either because of
the complexity of the structure of $c(u,\theta)$\cite{Bonnor},
or because the loss of mass does not occur for for a finite
number of terms in the expansion in
$n$\cite{Papapetrou2,Hallidy}. Bonnor and Rotenberg's
expression describes a radiative period between an initial
nonradiative, static state and a final nonradiative,
{\it nonstatic} state. The other approaches mentioned above
describe a radiative period between static initial and final
states.

Hobill\cite{Hobill} has provided an expression for
$c(u,\theta)$ that describes a radiative period between
initial and final static states, and that leads to a loss
of mass that is exactly equal to the total variation of the
mass aspect during the period. Again the expression is not
simple. It is given by

$$c(u,\mu)=
{{\lbrack m_b f(\mu)e^{\eta u}+m_a (1-\mu^2)\rbrack
(1+\eta \mu^2)(1-\mu^2)}\over
{f^2(\mu)e^{2\eta u}+(n-1)\mu^2-n\mu^4+1}}\;,\eqno(32)$$

\noindent where $n$ and $\eta$ are constants and $m_a$ is
identified as one fourth of the total mass loss. It relates
to the other constants according to

$${{18n}\over{n+1}}=\eta m_a\;,$$

$$m_b= \pm \sqrt{2\over n}\,m_a\;.$$

\noindent $f(\mu)$ is an arbitrary function that must be
everywhere regular and positive definite in the interval
$-1 \le \mu \le 1$. We note that $c$ vanishes over the
symmetry axis $(\mu=\pm 1)$. The news function obtained
from (32) reads

$${{\partial c}\over {\partial u}}=
{{\eta f(\mu)e^{\eta u}(1-\mu^2)}\over
{\lbrack e^{2\eta u}f^2(\mu)(1+n\mu^2)^{-1}+
(1-\mu^2)\rbrack^2}}
\biggl\{ {{-m_b f^2(\mu)e^{2\eta u}}\over
{1+n\mu^2}}$$

$$-{{2m_a(1-\mu^2)f(\mu)e^{\eta u}}\over
{1+n\mu^2}}+m_b(1-\mu^2) \biggr\} \;.\eqno(33)$$

\noindent Expressions (32) and (33) lead, via
integration of equation (28), to an extremely complicated
expression for the mass aspect:

$$M(u,\mu)=M(-\infty)-4m_a$$

$$+\biggl[ {{f^2(\mu)e^{2\eta u}}\over{1+n\mu^2}}
+(1-\mu^2)\biggr]^{-3}
\biggl[4m_a(1-\mu^2)^3+\sum^5_{l=1} H_l(\mu)e^{l\eta u}
\biggr]\;.\eqno(34)$$

\noindent The intricate expressions for $H_l(\mu)$ are given
in the Appendix of Ref. \cite{Hobill}. The problem with (32),
(33) and (34) is that they are too complicated to yield
numerical values for integrals (to be considered in the next
section) that contain these expressions, even by means of
computer calculations.

Therefore we attempted at obtaining a simpler expression
for $c(u,\mu)$. We damanded two conditions on $c(u,\mu)$.
First, it must satisfy the regularity conditions on the
$z$ axis, which guarantees that the system is permanently
isolated and leads to a well defined loss of mass. Second,
that the initial and final states are nonradiative. However,
as we will show, our expression leads to a nonstatic final
state. It is given by

$$c(u,\mu)={{ae^{nu}(1-\mu^2)F(\mu)}\over
{e^{2nu}+1}}\;,\eqno(35)$$

\noindent where $n$ and $a$ are constants ($n^{-1}$ and $a$
have dimension of length)
and $F(\mu)$ is a function that must be
choosen such that $c\,(1-\mu^2)^{-1}=c\,(sin\theta)^{-2}$ is
a regular function. Note that (35) vanishes for $\mu \pm 1$.
The news function associated to (35) is given by

$${{\partial c}\over{\partial u}}=
{{na(1-\mu^2)F(\mu)e^{nu}}\over
{e^{2nu}+1}}\biggl[1-
{{2e^{2nu}}\over{e^{2nu}+1}}\biggr]\;.\eqno(36)$$

\noindent In the limit $u\rightarrow \pm \infty$ we have

$$\lim_{u\rightarrow \pm \infty}c=
\lim_{u\rightarrow \pm \infty}{{\partial c}\over{\partial u}}
=0\;,\eqno(37)$$

\noindent for any nonvanishing value of $n$. The property
above is a necessary but not sufficient condition for having
a static final state. Equation (37) does not
determine whether the mass aspect $M$ depends on $\theta$,
and so from (29) there is the possibility that $d$ depends
on $u$. The initial state is assumed to be of the
Schwarzschild type.

Integrating equation (28) from $-\infty$ to $u$,
considering $c(u,\mu)$ given by (35), we obtain

$$M(u,\mu)=M(-\infty)-n^2a^2 (1-\mu^2)^2F^2(\mu)
\int_{-\infty}^u \biggl[{{e^{nu}}\over{e^{2nu}+1}}-
{{2e^{3nu}}\over{(e^{2nu}+1)^2}}\biggr]^2du$$

$$+{1\over 2}{{ae^{nu}}\over{(e^{2nu}+1)}}\biggl[
(1-\mu^2)^2 F'' -
8\mu(1-\mu^2) F'+4(3\mu^2-1)F(\mu)\biggr]
+\lambda(\mu)\;,\eqno(38)$$

\noindent where $F'={{dF}\over{d\mu}}$, 
$F''={{d^2F}\over{d\mu^2}}$ and $\lambda(\mu)$ is an
integrating function that depends only on $\mu$.

The integral in (38) depends on the sign of $n$. For $n<0$ we
have

$$M(u,\mu)=M(-\infty)+{1\over 6}na^2(1-\mu^2)^2F^2(\mu)
\biggl[{{3e^{4nu}+1}\over{(e^{2nu}+1)^3}}\biggr]$$

$$+{1\over 2}{{ae^{nu}}\over{(e^{2nu}+1)}}\biggl[
(1-\mu^2)^2 F'' -
8\mu(1-\mu^2) F'+4(3\mu^2-1)F(\mu)\biggr]
+\lambda(\mu)\;,\eqno(39)$$

\noindent and for $n>0$,

$$M(u,\mu)=M(-\infty)+na^2(1-\mu^2)^2F^2(\mu)
\biggl[ {1\over 6} {{3e^{4nu}+1}\over{(e^{2nu}+1)^3}}+
{{23}\over{48}}\biggr]$$

$$+{1\over 2}{{ae^{nu}}\over{(e^{2nu}+1)}}\biggl[
(1-\mu^2)^2 F'' -
8\mu(1-\mu^2) F'+4(3\mu^2-1)F(\mu)\biggr]
+\lambda(\mu)\;.\eqno(40)$$

By requiring the inital state to be static, expressions
(39) and (40) must satisfy

$$\lim_{u\rightarrow -\infty}M=
M(-\infty)\equiv M_0\;.\eqno(41)$$

\noindent Applying the condition above to (39) we find that
the function $\lambda(\mu)$ vanishes, and therefore

$$M(u,\mu)=M_0+{1\over 6}na^2(1-\mu^2)^2F^2(\mu)
\biggl[{{3e^{4nu}+1}\over{(e^{2nu}+1)^3}}\biggr]$$

$$+{1\over 2}{{ae^{nu}}\over{(e^{2nu}+1)}}\biggl[
(1-\mu^2)^2 F'' -
8\mu(1-\mu^2) F'+4(3\mu^2-1)F(\mu)\biggr]\;,\eqno(42)$$

\noindent for $n<0$. Similarly, applying (41) to
expression (40) we find that the integration function
$\lambda(\mu)$ is given by

$$\lambda(\mu)=-\biggl({1\over 6}+{{23}\over{48}}\biggr)
na^2 (1-\mu^2)F^2(\mu)\;.$$

\noindent Substituting it back in (40) for $n>0$ we
arrive at

$$M(u,\mu)=M_0+{1\over 6}na^2(1-\mu^2)^2F^2(\mu)
\biggl[{{3e^{4nu}+1}\over{(e^{2nu}+1)^3}}-1\biggr]$$

$$+{1\over 2}{{ae^{nu}}\over{(e^{2nu}+1)}}\biggl[
(1-\mu^2)^2 F'' -
8\mu(1-\mu^2) F'+4(3\mu^2-1)F(\mu)\biggr]\;.\eqno(43)$$

Let us now check the limiting value of (42) and (43) for
$u\rightarrow \infty$. Considering first (42) we obtain

$$\lim_{u\rightarrow\infty}M=M_0+
{1\over 6}na^2(1-\mu^2)^2F^2(\mu)\;,$$

\noindent where $n<0$. It follows from the expression above
that the total variation of the mass aspect $\Delta M_T$
associated with (42) is given by

$$\Delta M_T={1\over 6}na^2(1-\mu^2)^2F^2(\mu)\;.\eqno(44)$$

\noindent The mass aspect $M(u,\mu)$ given by (43), for
which $n>0$, leads to a similar expression for 
$\Delta M_T$:

$$\Delta M_T=-{1\over 6}na^2(1-\mu^2)^2F^2(\mu)\;.\eqno(45)$$

Hobill\cite{Hobill} obtained a result analogous to (44).
However his expression, in the limit $u \rightarrow\infty$,
does not depend on $\mu$. Consequently the final state in his
approach is static. In the present case, the mass aspect will
still depend on $\mu$ in the limit $u\rightarrow \infty$.
In view of (29) this fact implies  a final nonradiative,
nonstatic state, which according to Hobill may describe a
even more realistic situation.  In any event expressions (35)
and (36) describe an isolated physical system whose initial
and final states are nonradiative, and whose loss of mass is
precisely determined.\par
\bigskip
\bigskip
\noindent {\bf V. Evaluation of the gravitational energy}\par
\bigskip

In this section we  obtain numerical values for the
gravitational energy given by expression (9). The relevance of
such expression is that we can apply it to finite spacelike
volumes. In the present case we will calculate the
gravitational energy inside a large but finite surface  of
constant radius $r_0$, centered at the radiating source.
Expression (9) will be evaluated as a surface integral in the
asymptotic region where the components of Bondi's metric are
precisely determined. It reduces to just one integral given by

$$E\;=\;{1\over {8\pi}}\int_S d\theta\,d\phi\,eT^1\;,\eqno(46)$$

\noindent where $S$ is a surface of fixed radius $r_0$, assumed
to be large as compared with the dimension of the source, and
the determinant $e$ is given by $e=r^2\,A\,sin\theta$.

We will consider triads (13), (15) and
(25), for which the functions $c(u,\theta)$ and
$M(u,\theta)$ are given by (35) and (42), respectively. Thus
we take the mass aspect determined by the condition $n<0$,
only. Moreover we will consider four possibilities for the
function $F(\mu)$. Therefore we are effectively
analyzing twelve distinct sets of triads.

We will follow here the steps of section V of \cite{Maluf4}.
The energy expression that results from (13) and (15) have
already been evaluated. They are given by\cite{Maluf4}

$$E_1 \;=\;{ {r_0}\over 4}\int_0^\pi\,d\theta\,
\biggl\{ sin\theta\biggl[e^\gamma + e^{-\gamma}-
{2\over A}  \biggr]+
{1\over A}\, {{\partial}\over
{\partial \theta}}(Ur\,sin\theta)\biggr\}
\;,\eqno(47)$$

\noindent and

$$E_2\;=\;{r_0\over 4}\int_0^\pi d\theta\,
{1\over A}\biggl\{sin\theta
\biggl[ e^\gamma A'+e^{-2\gamma}A'B'-2+
e^{-2\gamma}{{\partial A'}\over {\partial \theta}}C'-BC'
-Be^{-\gamma}
{{\partial \gamma} \over {\partial \theta}}\biggr]$$

$$-B\,cos\theta\biggl[ B'-e^{-\gamma}\biggr]
\biggr\}\;,\eqno(48)$$

\noindent together with definitions (14) and (16). $E_1$ and
$E_2$ correspond to (13) and (15), respectively.
Unfortunately the sign of the first term in the expansion
of the function $U(u,r,\theta)$ of the metric tensor (10),
in Ref. \cite{Maluf4}, is changed. Therefore expressions
$E_1$ and $E_2$ given in the latter reference must be
corrected. Minor modifications (such as the modification of
some numerical coefficients) are
necessary. The correct expressions of $E_1$ and $E_2$ in
terms of $c(u,\theta)$ and $M(u,\theta)$ are given by

$$E_1\;=\;{1\over 2}\int_0^\pi d\theta\,sin\theta\,M
-{1\over {4r_0}}\int_0^\pi d\theta\,sin\theta
\biggl[ \biggl(
{{\partial c}\over {\partial \theta}} \biggr)^2\,
+4c\biggl({{\partial c}\over{\partial\theta}}\biggr)cot\theta
+4c^2\,cot^2\theta\biggr]$$

$$+{1\over {4r_0}}\int_0^\pi d\theta\,M\,
{\partial \over {\partial \theta}}
\biggl[sin\theta\biggl( {{\partial c}\over {\partial \theta}}
+2c\,cot\theta \biggr)\biggr]\;,\eqno(49)$$

$$E_2\;=\;{1\over 2}\int_0^\pi d\theta\,M\,sin\theta-
{1\over {4r_0}}\int_0^\pi d\theta\,sin\theta \biggl[
M^2+{1\over 2}\biggl(
{{\partial c}\over {\partial \theta}}\biggr)^2\,
+4c\biggl({{\partial c}\over {\partial \theta}}
\biggr)cot\theta$$

$$+6c^2cot^2\theta+\biggl(
{{\partial M}\over {\partial \theta}}\biggr)
\biggl({{\partial c}\over {\partial \theta}}+
2c\,cot\theta\biggr)
\biggr]
+{1\over {4r_0}}\int_0^\pi d\theta\,cos\theta\biggl[2c
\biggl({{\partial c}\over {\partial \theta}}\biggr)+
4c^2\,cot\theta\biggr]\;.\eqno(50)$$

\noindent It is clear that the total gravitational energy
given by both (49) and (50), in the limit
$r_0 \rightarrow \infty$, which corresponds to the limit
$u \rightarrow -\infty $, yield the same value, i.e., the
total mass:

$$\lim_{r\rightarrow \infty}E_1=
\lim_{r\rightarrow \infty}E_2=M(-\infty)
\equiv M_0\;.\eqno(51)$$

\noindent In fact it is proven in \cite{Maluf4} that the
total gravitational energies calculated out of triads
related by a local SO(3) transformation, and that have the
asymptotic behaviour given by (2), are the same.

We consider next triads given by (25). The components of the
torsion tensor are given by

$$\tilde T_{(1)12}=\biggl({c\over r}-{M\over r}+
r\partial_1({c\over r})+{1\over {2r}}\partial_2 f\biggr)
cos\theta\,cos\phi
-\biggl({1\over 2}\partial_1 f+{1\over {2r}}f+{1\over r}
\partial_2 M\biggr)sin\theta\,cos\phi\;,$$

$$\tilde T_{(1)13}=\biggl( {c\over r}+{M\over r}+
r\partial_1({c\over r})\biggr)sin\theta\,sin\phi-
{1\over {2r}}f\,cos\theta\,sin\phi\;,$$

$$\tilde T_{(1)23}=2c\,cos\theta\,sin\phi+
(\partial_2c-{1\over 2}f)sin\theta\,sin\phi\;,$$

$$\tilde T_{(2)12}=\biggl( {c\over r}-{M\over r}+r\partial_1
({c\over r})+{1\over {2r}}\partial_2 f \biggr)cos\theta\,
sin\phi
-\biggl( {1\over 2}\partial_1 f+{1\over 2}{f\over r}+
{1\over r}\partial_2M\biggr)sin\theta\,sin\phi\;,$$

$$\tilde T_{(2)13}=-\biggl( {c\over r}+{M\over r}+
r\partial_1({c\over r})\biggr)sin\theta\,cos\phi+
{1\over {2r}}f\,cos\theta\,cos\phi\;,$$

$$\tilde T_{(2)23}=-2c\,cos\theta\,cos\phi-
(\partial_2c-{1\over 2}f)sin\theta\,cos\phi\;,$$

$$\tilde T_{(3)12}=\biggl( {M\over r}-{c\over r}-
r\partial_1({c\over r})-{1\over {2r}}\partial_2 f\biggr)
sin\theta-\biggl({1\over 2}\partial_1f+{1\over {2r}}f+
{1\over r}\partial_2M\biggr)cos\theta\;,$$

$$\tilde T_{(3)13}=\tilde T_{(3)23}=0\;,\eqno(52)$$

\noindent where $\partial_1$ and $\partial_2$ denote partial
derivatives with respect to $r$ and $\theta$, respectively.

The calculation of (46) out of the components above does
not pose any particular problem, except that the calculation
is very long. Denoting by $\tilde E$ the gravitational energy
that follows from (25) and (52), we have

$$\tilde E=E_1-
{1\over {2r}}\biggl[ \int_0^\pi d\theta\,M^2sin\theta-
{1\over 4}\int_0^\pi\,d\theta
({{\partial c}\over{\partial\theta}}+
2c\,cotg\theta){{\partial c}\over{\partial\theta}}sin\theta$$

$$-{1\over 2}\int_0^\pi\,d\theta\,
({{\partial c}\over{\partial\theta}}+2c\,cotg\theta)
\,c\,cos\theta$$

$$+{1\over 8}\int_0^\pi\,d\theta
({{\partial c}\over{\partial \theta}}+2c\,cotg\theta)^2
sin\theta(3cos^2\theta-1)\biggr]\;.\eqno(53)$$

\noindent In the calculation above we have made use of
the regularity condition $c(u,\theta)=0$ for
$\theta=0,\pi$. As expected, (53) also satisfies (51).

The comparison of (49), (50) and (53) is crucial for the
selection of triads. We are now in a position  of obtaining
numerical values for these expressions. With this purpose
we make use of data based on the work of Saenz and
Shapiro\cite{Saenz} for assigning values to $M_0$ and $u$.
In the latter reference it is discussed a model for stellar
colapse of white-dwarves, with or without axial symmetry,
rotating or nonrotating. In this model gravitational waves
are produced during a burst of  $10^{-3}s$.
Thus $u$ may be taken to vary from $10^{-10}s$ to $10^{-4}s$.
Based on the work of Saenz and Shapiro we consider
a white-dwarf whose total mass is $M_0=1.4M_\odot$,
where $M_\odot$ is the solar mass. It is believed\cite{Ray}
that the maximum value for a white-dwarf is $1.4M_\odot$.
We will use geometrical unities (G=c=1) for the evaluation
of the energies. The value of $M_\odot$ in these unities is
$M_\odot=1.47664\times 10^5cm$. The distance $r$ will be
taken to vary from $10^{18}cm$ to $10^{13}cm$. Finally,
the parameters $a$ and $n$ assume the values $100$
and $-0.5$, respectively, in proper unities.

We have used the MAPLE V computer package to obtain $E_1$,
$E_2$ and $\tilde E$. For the function $F(\mu)$, which we
now denote $F(\theta)$, we take $sin\theta$, $sin^2\theta$,
$cos\theta$ and $cos^2\theta$. The resulting values are
listed in tables 1-4 for the particular value $u=10^{-4}s$.
The numerical values were obtained with a precision of
20 digits.

We have verified that for all these functions, for all
distances considered, $\tilde E$ is always the minimum
energy:

$$\tilde E < E_2 <E_1\;.$$

\noindent Note that since $E_1$, $E_2$ and
$\tilde E$ are calculated at constant $u=t-r$, the 
radial dependence of these expression is given just by the
${1\over {r_0}}$ coefficient in (49), (50) and (53). Thus
for constant $u$ the intricacy of these expressions reside
in the angular dependence.  We have further verified
numerically that the result above is also
obtained for any value of $u$ between $10^{-10}s$ and
$10^{-4}s$. The constants $a$ and $n$ must be chosen such
that the absolute values of
$\Delta M_T$ given by expressions (44) and (45) are not
greater than $M_0$, otherwise the total gravitational
energy will be negative. In any case,  at present
we do not know enough about axially-symmetric, nonrotating
isolated sources in order to provide realistic values for
these constants.\par

\begin{table}
\caption{$F(\theta)=sin\theta$}
\begin{center}
\begin{tabular}{|c||r|r|r|}\hline
    &  $r\sim 10^{18}cm$ & $r\sim 10^{15}cm$
    &$r\sim 10^{13}cm$ \\ \hline
    $E_1$ & 20653.912380952{\bf 37}8 &
            20653.9123809{\bf 52}3 &
            20653.9123{\bf 80}9 \\ \hline
   $E_2$ &  20653.912380952{\bf 36}1 &
            20653.9123809{\bf 35}0 &
            20653.9123{\bf 79}2  \\ \hline
$\tilde E$ &20653.912380952{\bf 34}4 &
            20653.9123809{\bf 17}8 &
            20653.9123{\bf 77}4 \\ \hline
\end{tabular}
\end{center}
\end{table}
\bigskip

\begin{table}
\caption{$F(\theta)=sin^2\theta$}
\begin{center}
\begin{tabular}{|c||r|r|r|}\hline
       & $r\sim 10^{18}cm$& $r\sim 10^{15}cm$
       & $r\sim 10^{13}cm$ \\ \hline
       $E_1$ & 20656.0287830687{\bf 80}9 &
               20656.0287830{\bf 68}78 &
               20656.0287{\bf 83}0 \\ \hline
       $E_2$ & 20656.0287830687{\bf 63}6 &
               20656.0287830{\bf 51}49 &
               20656.0287{\bf 81}3 \\ \hline
  $\tilde E$ & 20656.0287830687{\bf 46}3 &
               20656.0287830{\bf 34}21 &
               20656.0287{\bf 69}6 \\ \hline
\end{tabular}
\end{center}
\end{table}

\bigskip

\begin{table}
\caption{$F(\theta)=cos\theta$}
\begin{center}
\begin{tabular}{|c||r|r|r|}\hline
   & $r\sim 10^{18}cm$ & $r\sim 10^{15}cm$ &
   $r\sim 10^{13}cm$ \\ \hline
   $E_1$ & 20669.7853968253{\bf 96} &
           20669.785396{\bf 82}5 &
           20669.78539{\bf 68}3 \\ \hline
   $E_2$ & 20669.7853968253{\bf 79} &
           20669.785396{\bf 80}8 &
           20669.78539{\bf 50}9 \\ \hline
$\tilde E$&20669.7853968253{\bf 62} &
           20669.785396{\bf 79}0 &
           20669.78539{\bf 33}6 \\ \hline
\end{tabular}
\end{center}
\end{table}

\bigskip

\begin{table}
\caption{$F(\theta)=cos^2\theta$}
\begin{center}
\begin{tabular}{|c||r|r|r|}\hline
   & $r\sim 10^{18}cm$ & $r\sim 10^{15}cm$ &
   $r\sim 10^{13}cm$ \\ \hline
  $E_1$ &  20671.9017989417{\bf 99} &
           20671.9017989{\bf 41}8 &
           20671.90179{\bf 89}4 \\ \hline
  $E_2$ &  20671.9017989417{\bf 81} &
           20671.9017989{\bf 24}5 &
           20671.90179{\bf 72}1 \\ \hline
$\tilde E$&20671.9017989417{\bf 64} &
           20671.9017989{\bf 07}2 &
           20671.90179{\bf 54}8 \\ \hline
\end{tabular}
\end{center}
\end{table}

\bigskip
\noindent {\bf VI. Discussion}\par
\bigskip
The result of the previous section constitutes a strong
indication that the set of triads with asymptotic behaviour
given by equation (2), and that satisfies the symmetry condition
$h_{ij}=h_{ji}$, yields the minimum value for the gravitational
energy that is computed from expression (9). In addition to the
fact that the symmetrized triads are unique (namely, there does
not exist a second set of triads that satisfy (2) and the
symmetry condition), the present analysis indicates that the
correct description of the gravitational field in terms of
orthonormal triads, in the realm of the TEGR, is given by
(24). The results described above amount to an interesting
interplay between the energy properties of the space-time and
its tetrad description.

As long as we are interested in the dynamics of the
gravitational field only, as described by the metric tensor,
it is irrelevant which configuration of triads we adopt.
However, if we consider the dynamics of spinor fields, such
as the Dirac field, then the correct choice of triads is
crucial. Recall that the theory defined by (5) and (6) was
established under the asumption of the time gauge condition.
The latter, together with (25), establishes the complete set
of tetrad fields. And finally, if the detection of the
emission of gravitational energy carried by gravitational
waves is experimentally feasible, then the whole scheme
developed here will play a relevant role.

Bondi's radiating metric is valid only in the asymptotic region.
Therefore we can do no better than determining the asymptotic
behaviour of the triads. However in the more general case where
the metric tensor is valid everywhere, except for singularities,
we still have a prescription for assigning a unique set of
triads to a given metric tensor restricted to the spacelike
section. Such prescription is due to M\o ller\cite{Moller},
who called it ``supplementary conditions". Although he
established these conditions by still requiring de Donder
relations for the metric tensor, we can dispense with the
latter relations and ascribe generality to his proposal.
M\o ller suggested as supplementary conditions for the
space-time tetrad field the {\it weak field} condition

$$e_{a\mu} \approx \eta_{a\mu} +
 {1\over 2} h_{a\mu}\;,\eqno(54)$$

\noindent with $h_{a\mu}$ satisfying the symmetry condition
$h_{a\mu}=h_{\mu a}$. It differs from (2), which is restricted
to spacelike sections only, in that the symmetry condition
must be verified everywhere. Thus (54) is stronger than (2).
Nevertheless we can require the set of triads to satisfy (54)
in the general case. Every metric tensor for the spacelike
section of a space-time admits a unique set of triads that
satisfy a relation similar to (54), which is no longer a
boundary condition and therefore it can be applied to
space-times with arbitrary topology.
We finally mention  that the set of triads
presented in ref. \cite{Maluf3}, in the analysis of the
irreducible mass of a rotating black hole, satisfies  a
relation similar to (54), but in the three-dimensional
spacelike hypersurface.\par

\bigskip
\noindent {\it Acknowledgements}\par
\noindent K. H. C. B. is supported by CAPES, Brazil.\par

\end{document}